\documentclass[preprint,journal]{vgtc}       

\ifpdf
  \pdfoutput=1\relax                   
  \usepackage{graphicx}                
  \DeclareGraphicsExtensions{.pdf,.png,.jpg,.jpeg} 
\else
  \usepackage{graphicx}                
  \DeclareGraphicsExtensions{.eps}     
\fi%

\graphicspath{{figures/}{pictures/}{images/}{./}} 

\usepackage{microtype}                 
\PassOptionsToPackage{warn}{textcomp}  
\usepackage{textcomp}                  
\usepackage{mathptmx}                  
\usepackage{times}                     
\usepackage{cite}                      
\usepackage{tabu}                      
\usepackage{booktabs}                  
\usepackage{comment}
\usepackage{color}
\usepackage{fancyhdr}

\preprinttext{To appear in IEEE Transactions on Visualization and Computer Graphics.}

\onlineid{0}

\vgtccategory{Research}
\vgtcpapertype{please specify}

\title{Modulating Fine Roughness Perception of Vibrotactile \\Textured Surface using Pseudo-haptic Effect}

\author{Yusuke Ujitoko, \textit{Member, IEEE}, Yuki Ban \textit{Member, IEEE}, and Koichi Hirota \textit{Member, IEEE}}

\authorfooter{
\item
 Yusuke Ujitoko is with Research \& Development Group, Hitachi, Ltd., Yokohama, Japan. E-mail: yusuke.ujitoko.uz@hitachi.com.
\item
 Yuki Ban is with the Mechanical Engineering Department, the University of Tokyo, Chiba, Japan. E-mail: ban@edu.k.u-tokyo.ac.jp.
\item
 Koichi Hirota is with the Graduate School of Information Systems, The University of Electro-Communications, Tokyo, Japan. E-mail: hirota@vogue.is.uec.ac.jp
}

\shortauthortitle{Biv \MakeLowercase{\textit{et al.}}: Global Illumination for Fun and Profit}

\abstract{
Playing back vibrotactile signals through actuators is commonly used to simulate tactile feelings of virtual textured surfaces.
However, there is often a small mismatch between the simulated tactile feelings and intended tactile feelings by tactile designers.
Thus, a method of modulating the vibrotactile perception is required.
We focus on fine roughness perception and we propose a method using a pseudo-haptic effect to modulate fine roughness perception of vibrotactile texture.
Specifically, we visually modify the pointer's position on the screen slightly, which indicates the touch position on textured surfaces.
We hypothesized that if users receive vibrational feedback watching the pointer visually oscillating back/forth and left/right, users would believe the vibrotactile surfaces more uneven.
We also hypothesized that as the size of visual oscillation is getting larger, the amount of modification of roughness perception of vibrotactile surfaces would be larger.
We conducted user studies to test the hypotheses.
Results of first user study suggested that users felt vibrotactile texture with our method rougher than they did without our method at a high probability.
Results of second user study suggested that users felt different roughness for vibrational texture in response to the size of visual oscillation.
These results confirmed our hypotheses and they suggested that our method was effective.
Also, the same effect could potentially be applied to the visual movement of virtual hands or fingertips when users are interacting with virtual surfaces using their hands.
} 


\CCScatlist{ 
 \CCScat{K.6.1}{Management of Computing and Information Systems}%
{Project and People Management}{Life Cycle};
 \CCScat{K.7.m}{The Computing Profession}{Miscellaneous}{Ethics}
}

\teaser{
  \centering
  \includegraphics[width=5in]{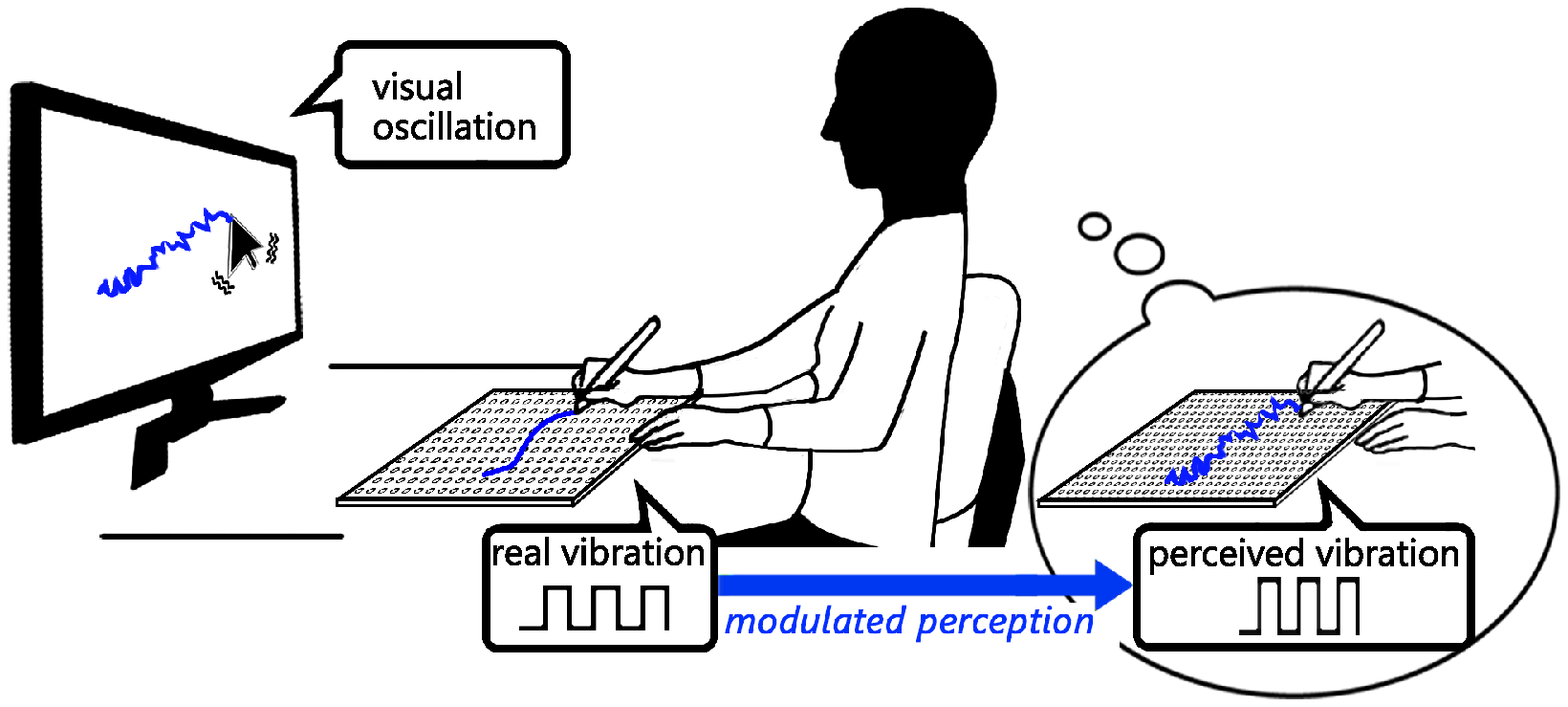}
  \caption{Our proposed method modulates the fine roughness perception of vibrotactile textured surfaces using pseudo-haptic effect. Our user study showed that users felt the surface rougher in response to a parameter configuration of visual feedback.}
	\label{fig_main_image}
}

\vgtcinsertpkg

\begin{document}
%
%


\firstsection{Introduction}
\maketitle

\thispagestyle{fancy}
\fancyhf{}
\lhead{ {\tiny © 2019 IEEE.  Personal use of this material is permitted.  Permission from IEEE must be obtained for all other uses, in any current or future media, including reprinting/republishing this material for advertising or promotional purposes, creating new collective works, for resale or redistribution to servers or lists, or reuse of any copyrighted component of this work in other works.\\}}


Due to the popularity of mobile devices with vibrotactile actuators, surface tactile technology has gained attention in recent years \cite{doi:10.1146/annurev-control-060117-105043}.
There are several approaches to presenting vibrations for interactions with virtual surfaces: vibrating the surfaces \cite{Hollins2000}, vibrating the tools \cite{Culbertson2017} (i.e., pens), or vibrating the skin directly by attached vibrators or indirectly by mid-air focused ultrasound beams\cite{Hasegawa2013,Hoshi:2010:NTD:1907654.1908041}.

Vibrotactile stimuli enable humans to perceive virtual textured surface properties.
In usual virtual reality (VR) applications, there are objects that have various virtual surfaces in virtual space, and users can touch them with their virtual bodies or indirectly using tools.
When vibrotactile stimuli that match textures of surfaces visualized in applications are presented to users, users perceive the surfaces as more realistic \cite{Culbertson2017}.
In order to realize this, developers must design a wide variety of vibrotactile feedback according to the texture configurations of surfaces.

One general design approach is to collect recorded vibrotactile signals that match the contents or the scenes of the applications and apply the signals there.
It has become easier to collect recorded signals recently because of off-the-shelf vibrotactile signal recording devices \cite{Minamizawa2012} and public vibrotactile signal datasets \cite{Strese2017}.
Developers now have these means to obtain recorded signals.

However, it is difficult to convey the intended surface tactile feeling with recorded signals.
There is often a slight mismatch between what tactile feeling recorded signals convey and what tactile feeling developers intend.
Some factors that contribute to a mismatch are as follows.
There is a possibility that recorded signals contain some noise.
There is a possibility that the recorded environments may be somewhat different from the assumed scenes in the applications.
There is a possibility that the replaying devices can vary the recorded signal output to the users due to the devices'{} limited specification \cite{doi:10.1080/00140139408964917, 5722959}.
Also, there is a possibility that the mechanical coupling between a vibrotactile display and the skin during replaying is different from that during recorded \cite{doi:10.1121/1.2715669}.
All of the above factors are related to signal collection and signal display.
When developers notice a mismatch, they must modulate the vibrotactile feeling so that users feel the surface information as intended.
We propose a novel method for modulating the vibrotactile perception of texture.
Specifically, we focus on fine roughness as a target, which is one of the important texture perceptual dimensions.

Previous studies have proposed approaches to modulating fine roughness perception.
These studies attempted to stimulate users{}' fingers by additional vibration.
We refer to this approach as a "signal-based approach."
Hollins et al. \cite{Hollins2000} showed that the vibrating surface was judged by users to be smoother less often than surfaces where no vibration was presented.
Asano et al. \cite{Asano2015} introduced a method of applying additional vibrational stimuli on users'{} fingers to modify roughness perception.
This signal-based approach made full use of vibrotactile signals to modify perception, and it succeeded in displaying rougher textured surface.
A disadvantage of this approach is that it could be affected by factors related to signal display such as vibrational device specifications.
In contrast to the signal-based approach, we aim at modulating the perceptual fine roughness of vibrotactile textures with a pseudo-haptic effect.
We categorize our approach as a "cross-modal approach."
The cross-modal approach makes full use of visual stimuli to modify the perception of displayed signals.
The cross-modal approach can be applied to the any displayed signal, even if it is changed by device specification.

\begin{figure}[h]
\centering
\includegraphics[width=3.0in]{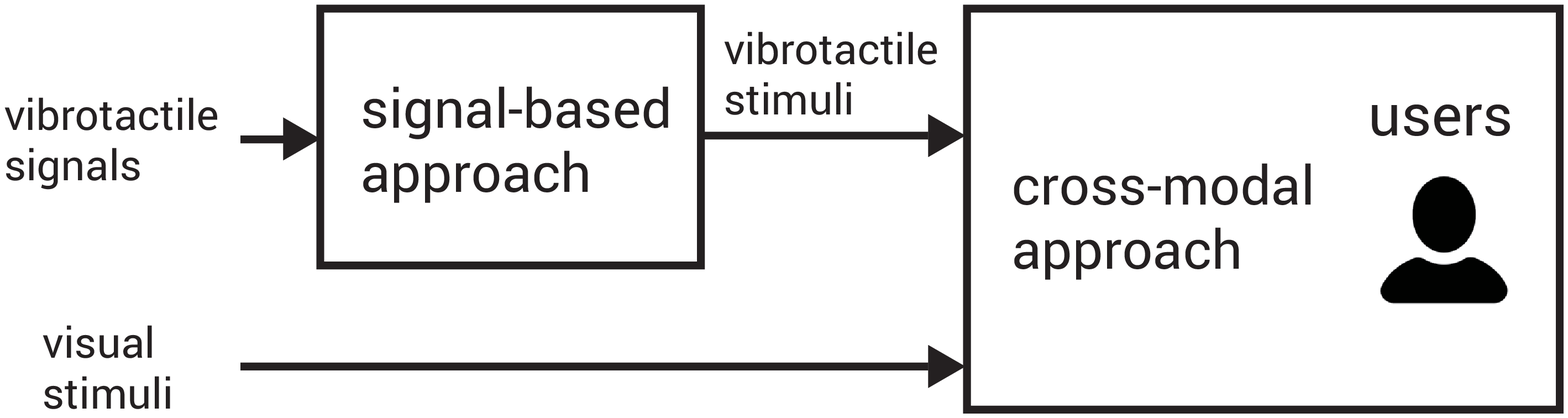}
\caption{The signal-based approach and our cross-modal approach can be used together by stacking them.}
\label{fig_stack}
\end{figure}

The signal-based approach takes a vibrotactile signal as input and outputs stimuli to users.
The cross-modal approach takes vibrotactile and visual stimuli as inputs to display pseudo-haptics to users.
Thus, our cross-modal approach can be used in combination with the signal-based approach by stacking both approaches in series (Fig.\ref{fig_stack}).
In this way, the cross-modal approach plays a key role in fine-tuning roughness perception.
Here, we will first clarify the effectiveness of our cross-modal approach when it is used on its own.

Recently, studies are increasingly focusing on pseudo-haptics.
Pseudo-haptics is a representative of a cross-modal effect between visual and haptic senses \cite{Lecuyer2009,Issartel2015PerceivingMI,Jauregui:2014:TPA:2608067.2608146}.
The pseudo-haptic effect indicates the haptic perception evoked by vision in haptic sensations.
The sensation is produced by an appropriate sensory inconsistency between the physical movement of the body and the observed movement of a virtual pointer.
For example, when a pointer decelerates in a standard desktop environment with a mouse and a pointer, users feel a sense of friction, gravity, or viscosity without any haptic actuator \cite{Lecuyer2000}.

Current pseudo-haptic studies have demonstrated cases where the pseudo-haptic effect modifies the perception of real haptic feedback.
Ban et al. \cite{banCurvature} showed that pseudo-haptics altered users'{} proprioceptive sensations corresponding to visual sensations by the combination of visual and haptic sensations.
They tried to modify the perception of curvature \cite{banCurvature}, angle \cite{banAngle}, and size \cite{banSize} of an object by presenting both distorted visual information and an actual object as a haptic cue.
Though these studies suggest that pseudo-haptics'{} capability to modulate perception of real haptic feedback, there are no studies on modulating the perception of real vibrotactile texture using pseudo-haptic effect.

As a result, in this study, we attempted to modulate vibrotactile surface perception using pseudo-haptic effect.
Specifically, surface fine roughness, which is an important dimension in the tactile definition of texture, is our target of modulation.
We hypothesized that if users watch the pointer slightly oscillating back/forth and left/right while receiving vibrational feedback, they would believe that the vibrotactile surface is rougher.
Our hypotheses are described in full in Section 3.
We quantitatively evaluated the hypotheses in two user studies.

We hope that our method can be applied to the head-mounted display-based VR system in the future.
However, we conducted an evaluation as a first attempt in these studies using a desktop environments with input touch device and screen based system where there are many previous studies succeeded in presenting pseudo-haptics.
We presented vibration through vibrating pen-type touch device because they are low cost and simple to implement.
The results of the user study showed the effectiveness of our method to modulate the fine roughness of textured surfaces.
Our method can be applied if only there is a virtual pointer on surfaces and vibrotactile feedback.
Thus, when VR developers would like to use our method in order to modulate the perception of vibrational roughness, they only need to visualize the pointer indicating the position of touch on surfaces and simultaneously present vibration to users in some way.
For the VR applications, we can freely visualize the pointer at the contact area on the surface as opposed to the touchscreen applications and thus, our method can be used easily to VR applications.
We hope that our method will help developers to overcome mismatches and will promote the rapid design of vibrational applications.
The contributions of this study are as follows:

\begin{itemize}
 \item We proposed a new method of modulating fine roughness perception of vibrotactile textured surfaces using pseudo-haptic effect. We regard our method as a cross-modal approach, which is in contrast to conventional signal-based approaches such as the ones by \cite{Culbertson2012, Asano2015}. It is expected that both approaches are compatible.
 \item A user study showed that users felt vibrotactile texture as significantly rougher with our method than they did without our method. We used a square wave signal as a patterned texture and we confirmed that our method was effective with all variations of the frequency of square wave that were used in the user study.
 \item Another user study quantitatively evaluated the change in roughness perception with our method. Users felt as if the amplitude of the presented square wave grew as the size of visual oscillation became larger. These results suggest that it is possible to modulate vibrotactile roughness perception by configuring of our method.
\end{itemize}

The remainder of this paper is organized as follows.
In the next section, we describe related work.
Afterwards, we present the concept of our work and introduce our research questions.
We then present two user studies. Finally, we conclude our paper.

\section{Related work}

Vibration, as a low-cost substitute for force feedback, is commonly used to simulate tactile feelings of textures for interactive devices and is proven to be an effective means of conveying tactile information \cite{Tashiro2009, Maeno2006, Strohmeier:2017:GHT:3025453.3025812}.
Vibration feedback is currently available in off-the-shelf consumer devices such as tablet computers and smartphones.
Developers can provide users with enjoyable tactile applications with such devices.
However, developers have difficulty modulating vibrotactile experiences.
Though there are recording devices for vibrotactile signals \cite{Minamizawa2012} and vibrotactile datasets \cite{Strese2017}, such vibrotactile signals cannot be applied to specific scenes or contents in applications without additional tuning of them, so modulation of them is required

\subsection{Signal-based modulation of vibrotactile feelings}

Previous studies have applied additional vibrotactile signals to users and modulated fine roughness perception.
Hollins et al. \cite{Hollins2000} presented users with two surfaces, one stationary and one vibrating.
They found that users tended to perceive the vibrating surface as rougher than the stationary one.
Asano et al. \cite{Asano2012} proposed a method of selectively modifying the roughness sensations of real materials by applying additional vibrotactile stimuli to users'{} finger pads.
They verified through user studies that their method successfully modulated roughness perception.

In contrast to the signal-based approach, we use a cross-modal display method that is able to modulate the perceptual fine roughness of texture.
We categorize our approach as a "cross-modal approach," which solves problems by resorting to cross-modal visuo-haptic interaction.
Signal-based approaches and our approach are compatible.
It is expected that the amount of modulation of perceptual roughness would be larger when both approaches are used together.

\subsection{Cross-modal modulation of vibrotactile feelings}

Cross-modal visuo-haptic interaction has been studied for a long time.
Studies on it are based on the key idea that when visual and other senses conflict, vision often dominates in multisensory integration, so sensory input can be distorted in favor of vision.
For example, in their classic experiment, Rock and Victor \cite{Rock1964} asked users to look at and touch an object.
They created a conflict between vision and touch by distorting the visually perceived shape from the actual shape perceived by touch.
As a result, users reported that the object felt the way it looked, suggesting that the conflict between vision and touch was completely resolved in favor of vision, and users were unaware of the conflict.

\subsubsection{Modification of texture appearance}

As for texture perception, Lederman et al.\cite{Lederman1981} and Heller \cite{Heller1982} showed that vision and haptics performed equally well in texture perception tasks.
Their results showed that the discrepancy between the roughness of visual and tactile senses led to an equal weighting of information from both senses.
Based on their results, one may consider a method of controlling visual roughness of textured surfaces to modulate tactile perceptual roughness.
However, in most cases, the appearance of textured surfaces cannot be modified, and a method that modifies texture appearance cannot be applied to practical applications such as smartphone games.
Thus, a method that does not affect the visual information of a textured surface is required.

\subsubsection{Modification of pointer's movement}
Recently, there have been seminal works that focus on pseudo-haptics, which makes full use of cross-modal effect to render haptic feelings.
Pseudo-haptic sensation occurs when physical body movement differs from the observed movement of a virtual pointer on screen \cite{Lecuyer2009}.
When a user believes that the pointer moves according to the movement of their body, changes in the movement of the pointer are regarded as changes in the haptic sense, such as force or friction on the hands, and evoke a pseudo-haptic sensation.

As for texture perception, there have been studies that attempt to generate texture perception using pseudo-haptic effects without any haptic device.
The target surface perception of these studies has been macro roughness \cite{Lecuyer2004}, fine roughness \cite{Watanabe2008}, and stiffness \cite{Argelaguet:2013:EIP:2506206.2501599, Hachisu2011}.

The work in \cite{Lecuyer2004} showed that users could successfully identify macroscopic textures such as bumps and holes by simply using variations in the motion of the pointer without any haptic device.
They adjusted the motion of the pointer as a function of the simulated height of the macroscopic textures over which the pointer is traveling.
Users were able to draw the different profiles of simulated bumps and holes correctly.

The work of Watanabe et al. \cite{Watanabe2008} proposed a distorted pointer visualization system, where the pointer position changed at random by 2-6 pixels while the mouse was moved.
Their objective was to communicate tactile impression via only visual oscillation of the pointer.
They claimed that their method generated virtual roughness perception without any haptic device.
However, they did not conduct user studies, so it is unclear whether their method was effective.

Argelaguet et al. \cite{Argelaguet:2013:EIP:2506206.2501599} showed that visual feedback was able to induce a sensation of stiffness when the user interacted with an image using a standard mouse without any haptic device.
Once the user clicked on the image, the deformation was driven by the stiffness coefficient of the image and the time the user kept the mouse button pressed.
The results from the user study showed that users were able to efficiently distinguish variations of the stiffness coefficient by up to 14\%.

All of the works presented above visually distorted pointer's movement to induce the pseudo-haptic effect.
Their objective was to communicate with users{}' tactile texture perception via only visual information.
Thus, all of the works assumed that the environments had no real haptic information.
However, as the previous study \cite{Tatezono2009} showed, the combination of pseudo-haptic feedback and real haptic feedback makes users feel stronger perceptual force than when there is only pseudo-haptic feedback.

In this study, we use the pseudo-haptic effect to modulate the roughness perception of vibrotactile textures
Work by Hachisu et al. \cite{Hachisu2011} had a similar objective to ours.
They presented a method of modulating the perceived stiffness of a real object during tapping.
They changed the pointer visualization parameter of collision between the stick and keyboard instruments.
Their method made users feel rich haptic sensations.
Though their study had similar objective to ours, the interaction was different from ours.
Their interaction was tapping on the keyboard instruments and ours are scanning the textured surface.
In addition, their target dimension was stiffness while ours is roughness.

\section{Proposed method}
\subsection{Concept and hypotheses}

The objective of this study is to modulate the fine roughness perception of vibrotactile textured surfaces while users are scanning the surfaces and receiving vibrational feedback.
This study uses the pseudo-haptic effect to realize this because it has been shown to be a simple approach in many previous studies \cite{Lecuyer2009,Lecuyer2004}.
It has been shown that an appropriate visuo-haptic inconsistency between the movement of the user's input and the observed movement of a virtual pointer could lead to pseudo-haptic feelings.
Thus, in order to induce the pseudo-haptic effect, we must visually distort the virtual pointer in some way.

\begin{figure}[h]
\centering
\includegraphics[width=3.5in]{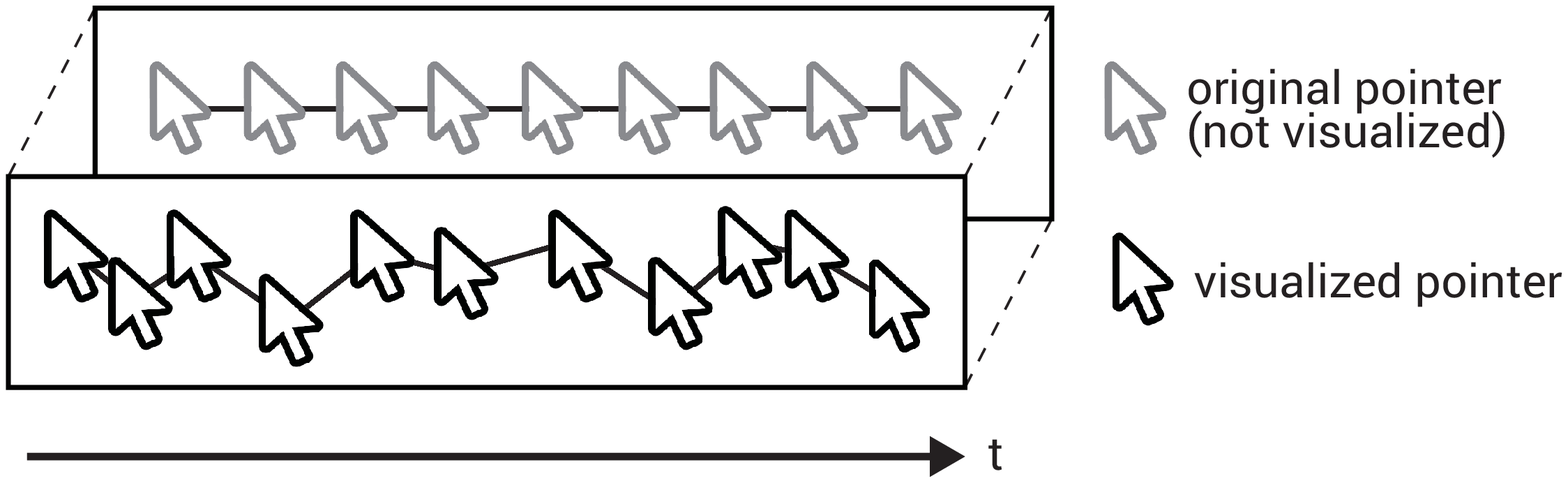}
\caption{One example of conditions where pseudo-haptic effect occurs. There is a difference between the trajectories of the virtual pointer, which oscillates back/forth and left/right, and the original pointer.}
\label{fig_concept_osc}
\end{figure}

In the real world, when users scan rough surfaces such as sandpaper with a pen, they observe the point of touch visually oscillating slightly as a result of the pen-surface physical interaction.
Thus, we hypothesize that if users watch the pointer overly oscillating back/forth and left/right while exploring the textured surfaces and receiving vibrational feedback, users would see the vibrotactile surfaces as more uneven.
Fig.\ref{fig_concept_osc} illustrates this concept.
In this illustration, users attempt to move the pointer along the horizontal axis, but the visualized pointer's movement is slightly translated in a random direction by the system.

The reason why we do not adopt the pointer's behavior of just momentarily stopping in its forward movement is that it seems closer to the behavior of an object encountering friction.
The friction and roughness is different texture perceptual dimension and we focus on modulating roughness rather than friction perception in this study.

\begin{figure}[h]
\centering
\includegraphics[width=3.5in]{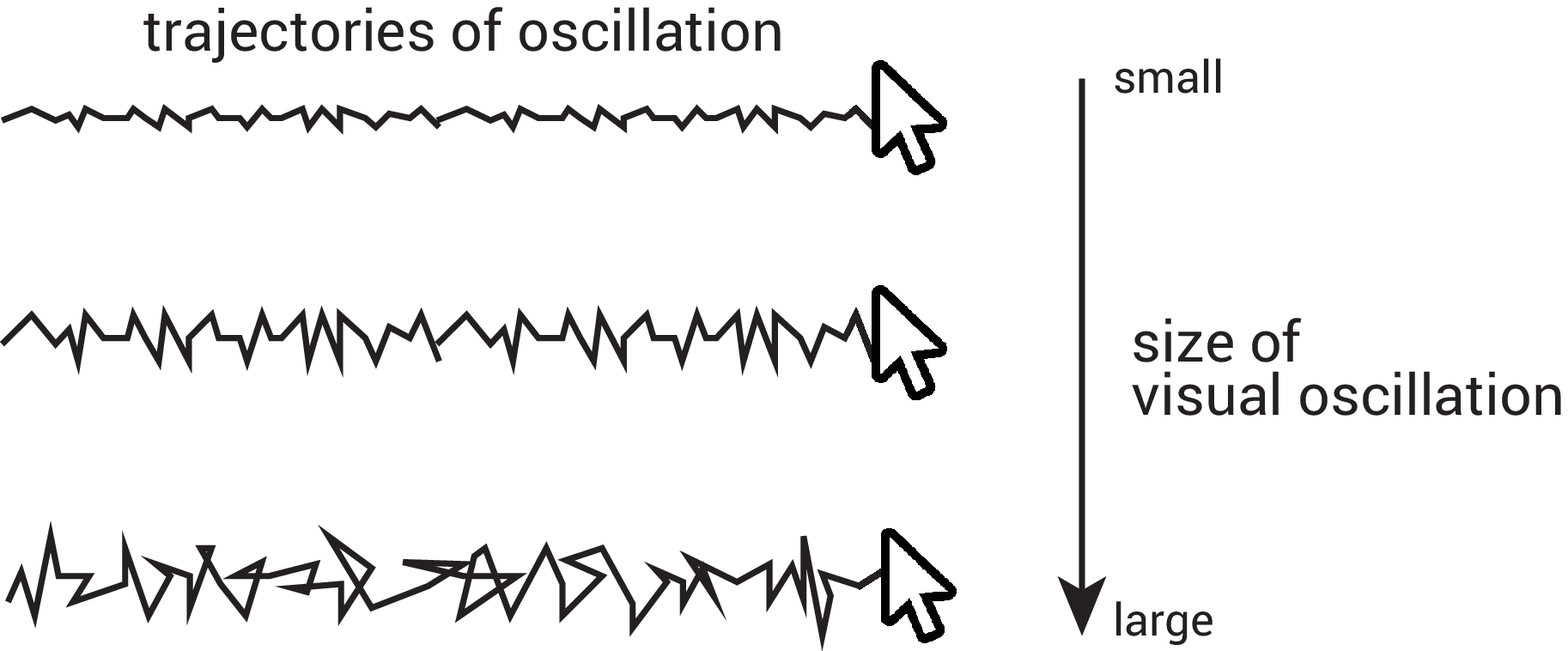}
\caption{Trajectories of the visual oscillation according to the configuration of oscillation size.}
\label{fig_concept_size_of_osc}
\end{figure}

We also hypothesize that the size of visual oscillation would affect the amount of modulation of roughness perception.
For example, when the size of the visual oscillation is small, the amount of modulation would be small.
In contrast, when the size of the oscillation is large, the amount of modulation would be large.
Fig.\ref{fig_concept_size_of_osc} illustrates the trajectories of the visual oscillation when the configuration of oscillation size is different.

The hypotheses stated in this section are summarized as follows:

\begin{description}
 \item[H1] If users watch the pointer visually oscillating back/forth and left/right while receiving vibrational feedback, they believe that the vibrotactile surfaces are rougher.
 \item[H2] As the size of visual oscillation becomes larger, the amount of modification of roughness perception of vibrotactile surfaces would be larger.

\end{description}

To test the hypotheses, we conducted two user studies.
User study 1 in Section 4 tested H1, and user study 2 in Section 5 tested H2.

\subsection{Implementation}

\begin{figure}[h]
\centering
\includegraphics[width=3.0in]{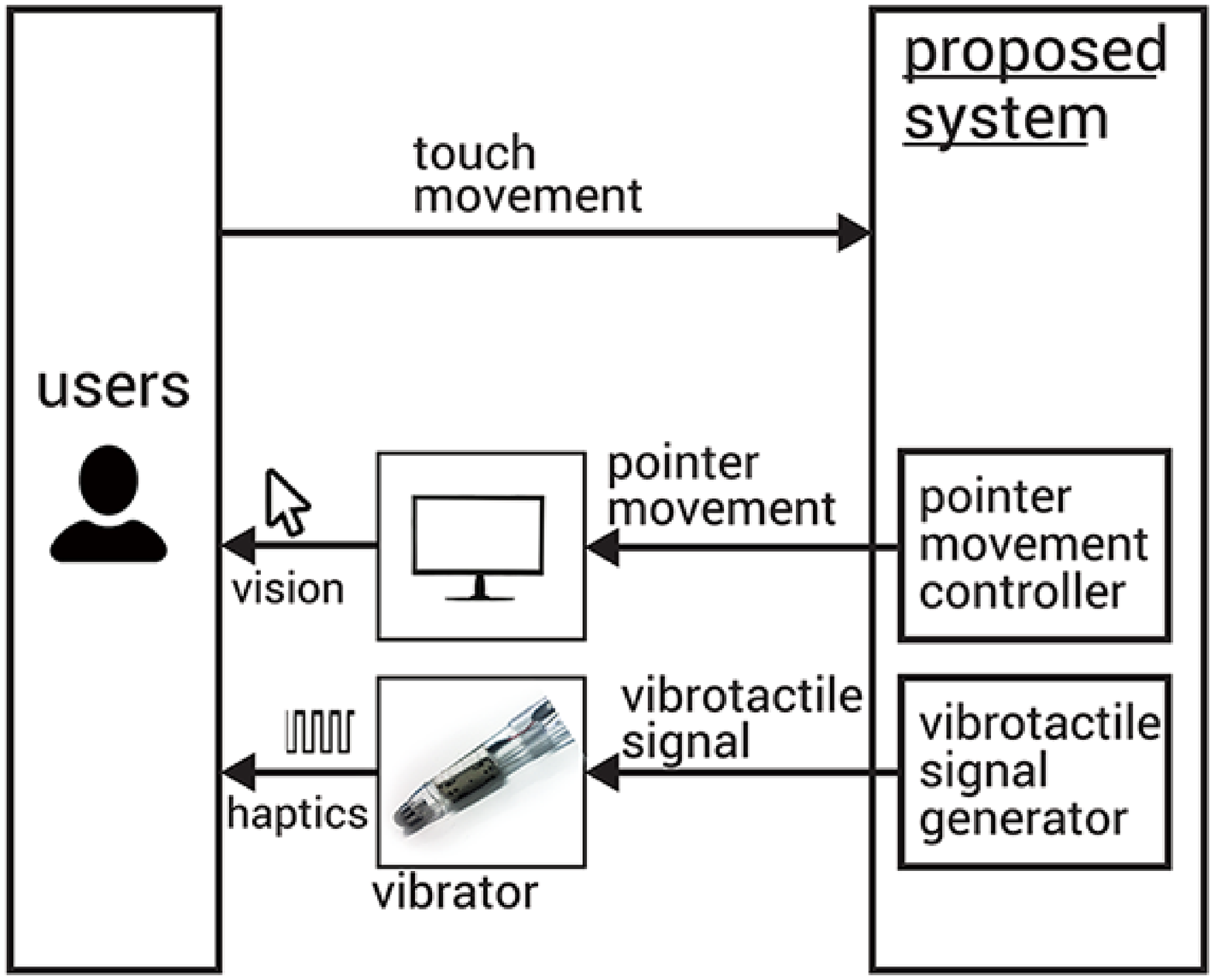}
\caption{Data flow during interaction between users and the system. Users input touch information to the system. The proposed system provides users with distorted pointer movement as visual and vibrotactile information.}
\label{fig_data_flow}
\end{figure}

We implemented a simple system that realizes the concept.
We conduct user studies using the system described here.
The data flow during interaction between users and proposed systems is illustrated in fig.\ref{fig_data_flow}.

We assume the use cases where users explore the surface textures using a pen-type device.
Users hold the pen-type device in which the vibrator is embedded.
Users move the device on touch interfaces such as touchpads.
The touch information such as touch timing and position on the touchpad is transmitted to the proposed system when users move the device.
The system updates the user touch information and visualizes the pointer and output the vibrotactile signals in return.

The system has two key functional modules: a pointer movement controller and a vibrotactile signal generator.
The pointer movement controller plays a role in deciding the pointer movement based on the user input touch positions.
The ordinary system does not need to modify the pointer position and it returns the position as users move the pen.
In contrast, our system manipulates the pointer position based on the user input.
How our system manipulates the pointer position is described in the next paragraph.
The vibrotactile signal generator plays a role in providing vibrotactile signals in some way.
There could be many ways to generate signals.
In our user studies, we generate a square wave to present patterned textured surfaces.
Through the pen-type device, users feel the vibration of the virtual textured surface during the pen-surface interaction.
The perception of vibration is the target of modulation in this study.

We translate the pointer position at random directions when the touch position is updated.
The pointer translation is calculated as

\begin{eqnarray}
{X}_{vis} &=& {X}_{origin} + {X}_{delta} \\
{Y}_{vis} &=& {Y}_{origin} + {Y}_{delta} \\
{X}_{delta} &=& C * \alpha * random(-1, 1) * abs(V) \\
{Y}_{delta} &=& C * \alpha * random(-1, 1) * abs(V)
\end{eqnarray}

\begin{figure}[h]
\centering
\includegraphics[width=3.0in]{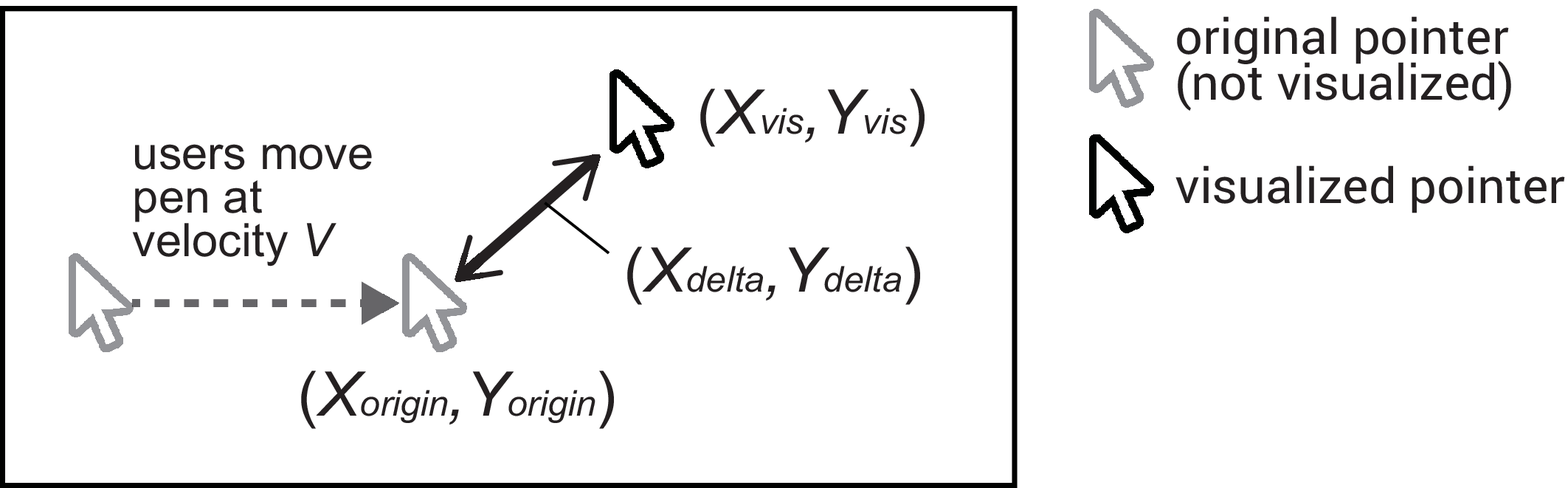}
\caption{The original pointer and visualized pointer. The visualized pointer's position is decided based on the original pointer's position and velocity.}
\label{fig_algorithm}
\end{figure}

Here, ${X}_{origin}$ is the original pointer position along the x-axis and ${Y}_{origin}$ is the original pointer position along the y-axis.
The x-axis and the y-axis are the horizontal and vertical axis on the screen, respectively.
${X}_{origin}$ and ${Y}_{origin}$ are the coordinates of the point that users intend the pointer be.
${X}_{vis}$ is translated pointer position along the x-axis.
${Y}_{vis}$ is translated pointer position along the y-axis.
${X}_{vis}$ and ${Y}_{vis}$ are the respective x and y coordinates of the point the users watch on the screen.
${X}_{delta}$ and ${Y}_{delta}$ are the amount of translation and they are derived by multiplying between $\alpha$, a random value from the uniform distribution (-1, 1), and the absolute value of the device's velocity $V$.
The reason the velocity $V$ is multiplied is that visual roughness is proportional to the number of points of patterned unevenness on the surface per unit scanning time.
Every time our system captures the user's pen device's position, it calculates ${X}_{vis}$ and ${Y}_{vis}$ and visualizes the virtual pointer on the screen.
$C$ is a constant value.
$\alpha$ is the coefficient unit, which defines the "size of visual oscillation."
As $\alpha$ or the device's velocity $V$ becomes larger, the amount of translation becomes larger.
The relationship among these parameters is visualized in Fig.\ref{fig_algorithm}.

\section{User study 1}

User study 1 was conducted to test hypothesis H1.
In other words, we tested whether participants feel vibrotactile surfaces as rougher when they watch a visible pointer oscillating.

We briefly introduce the study design and the variables we manipulated.
We performed a within-participants study relying on a 3x4 factorial design crossing four visual conditions and three vibration conditions.
We hypothesized that the visual conditions variables would have an effect on perceived roughness.

There were ten participants (eight males and two females) aged from 22 to 25.
All of the participants were right-handed.
They were screened to determine that they were not depressed or tired because perception can be affected by physical or emotional states.
The University of Tokyo Ethics Committee approved the data acquisition in this paper and written informed consent was obtained from all participants.

\subsection{Experimental system}

\begin{figure}[h]
\centering
\includegraphics[width=3.5in]{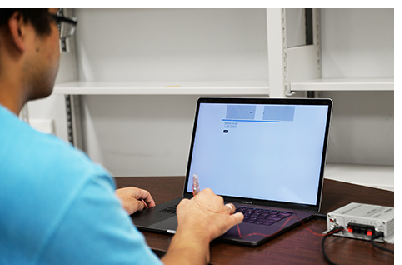}
\caption{Experimental system. The task in this user study was to move a pen-type device on a touchpad surface while receiving vibrotactile feedback via the pen}
\label{fig_ex1_experiment}
\end{figure}

The task in this user study was to move a pen-type device on a touchpad surface while receiving vibrotactile feedback via the pen (Fig.\ref{fig_ex1_experiment}).
By moving the pen-type device, participants operated a virtual pointer on two different rectangular areas on a screen.
Participants compared the roughness of texture between two areas where the configuration of vibrational feedback was exactly the same between them, but the visual oscillation was enabled under one condition and was disabled under the other condition.

\subsubsection{Experimental pen-devices}
\begin{figure}[h]
\centering
\includegraphics[width=3.5in]{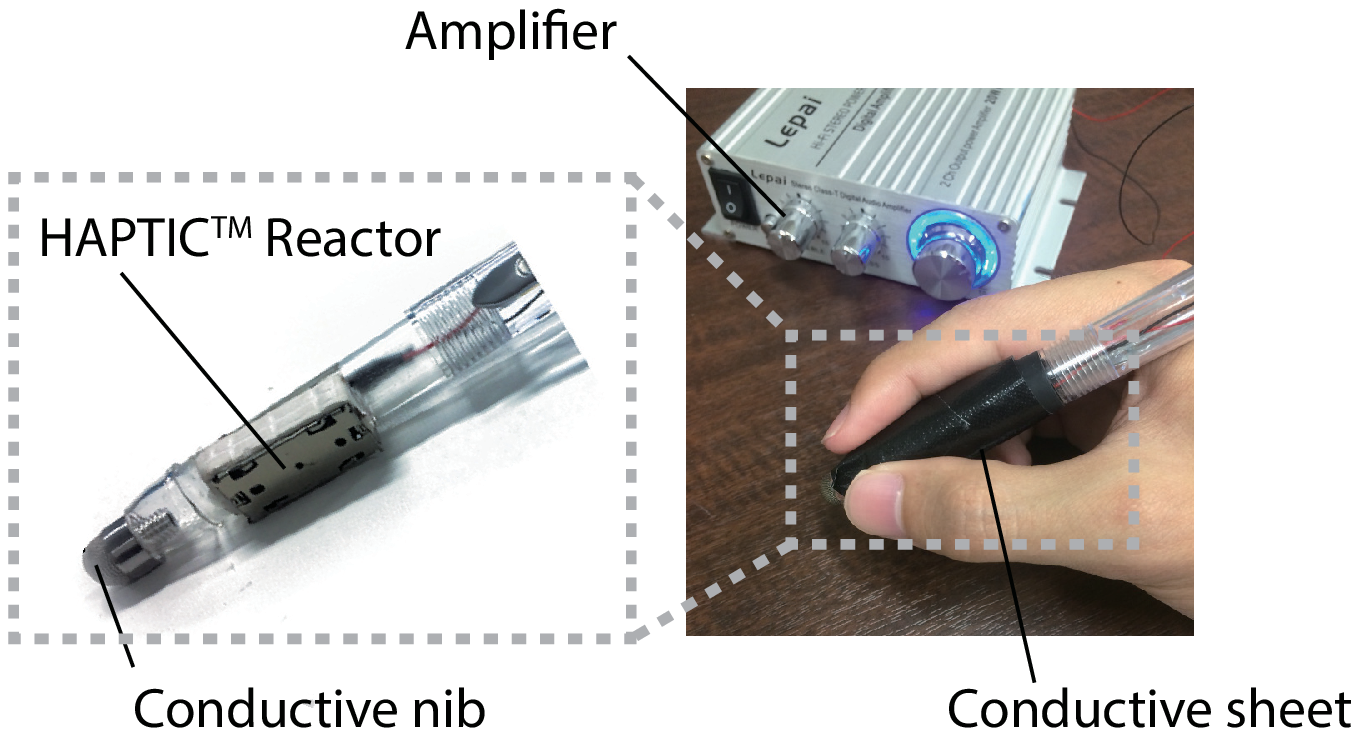}
\caption{Experimental vibrating pen device.}
\label{fig_ex1_pen}
\end{figure}

Our experimental system was composed of a laptop PC (Apple Inc., Mac Book Pro 15 inch, 220ppi) with a touchpad and a 2880 x 1800 pixels retinal display, a signal amplifier (Lepai Inc., LP-2020A ), and a vibrating pen-type device (Fig.\ref{fig_ex1_pen}).
The pen-type device, which we handcrafted, is described in more detail in the next paragraph.

The pen device was approximately 140 mm in length and weighed approximately 20 g.
The diameter of the grip part of the pen was approximately 10 mm.
We covered the pen tip with a conductive mater ial because the shaft of the pen was plastic and did not conduct to the grip part.
We wound a conductive sheet onto the grip to react with a capacitance type touch screen.
We embedded a vibrator (ALPS Inc., HAPTIC Reactor) inside the pen-type device 2 cm from the tip of the pen, where participants gripped.
The vibrator was small (35.0 mm $\times$ 5.0 mm $\times$ 7.5 mm) and light (approximately 5 g), so participants did not tire when moving the pen.
When participants moved the pen on the touchpad, the vibration signal was emitted from the earphone jack of the laptop.
The amplifier amplified the signal and the vibrator embedded in the pen presented the vibration to the participants'{} fingers.

\subsubsection{Configuration of pointer movement controller}
The system manipulated the pointer position as described in Section 3.
The parameter $\alpha$, which affects the size of visual oscillation, was a variable in this experiment.
In this experiment, we set $\alpha$ as 0, $1.5$, $2$, $2.5$, or $3$.
$\alpha$ on one of the two rectangular areas was set to 0. We refer to this area as a "non-oscillatory area."
$\alpha$ on the other area was assigned as $1.5$, $2$, $2.5$, or $3$. We refer to this area as an "oscillatory area."
Also, we refer to the conditions of the oscillatory area with different $\alpha$ as "visual conditions."

We conducted informal evaluations to define the parameters so that the vibrations presented to participants'{} fingers were neither too large nor too small.
In this experiment, we set $C$ so that when users move the pen at a speed of 10.4 mm/s (90 pixels/s) and $\alpha$ is set as $2$, ${X}_{delta}$ and ${Y}_{delta}$ are uniformly sampled from -0.21 mm (1.8 pixels) to 0.21 mm (1.8 pixels).
When $\alpha$ is set as 3, ${X}_{delta}$ and ${Y}_{delta}$ are uniformly sampled from -0.3 mm (2.7 pixels) to 0.3mm (2.7 pixels).

Note that in the implemented environments used in the user studies, only integer values were supported.
Thus, when there is a floating point, pixel position was either rounded down or placed in the next cell.
For example, 1.2 becomes 1 and 1.7 becomes 2.

\subsubsection{Configuration of vibrotactile signal generator}

As a first step to investigate the pseudo-haptic effect on vibrotactile perception, we assumed a simple patterned virtual surface to test the effect of visual oscillation.
Therefore, we used the simplest waveform to present vibrations to users, the wavelength and amplitude of which is easily controllable.
The waveforms that meet our requirements include square, sine, triangle, etc., but we chose to use a square wave.
The virtual textured surface is enhanced with a striped pattern that generates a square wave vibration feedback of constant wavelength $\lambda$:

\begin{equation}
y(t) = A\; sgn(\sin (2 \pi \frac{V(t)}{\lambda} + \varphi))
\end{equation}

The frequency of the vibrotactile signal depends on the velocity $V$ of the pen.
In this experiment, we set $\lambda$ as 1/3, 1/5, or 1/7.
For example, when $\lambda$ was set as 1/5 and the pen's movement speed was 10.4 mm/s (90 pixels/s), the frequency of the square wave was 9 Hz and the wavelength was 1.15 mm (10 pixels).
We call the vibrotactile signal variable "signal conditions" in this user study.

$A$ represents amplitude, and $\varphi$ represents the phase of the square wave.
In this experiment, $A$ was constant.
$A$ affected the applied voltage on the vibrator.
We measured the peak-to-peak voltage on the vibrator when $\lambda$ was 1/5 and the pen's movement speed was 10.4 mm/s. The voltage was 4.67 V.

\subsection{Task design}

\begin{figure}[h]
\centering
\includegraphics[width=3.5in]{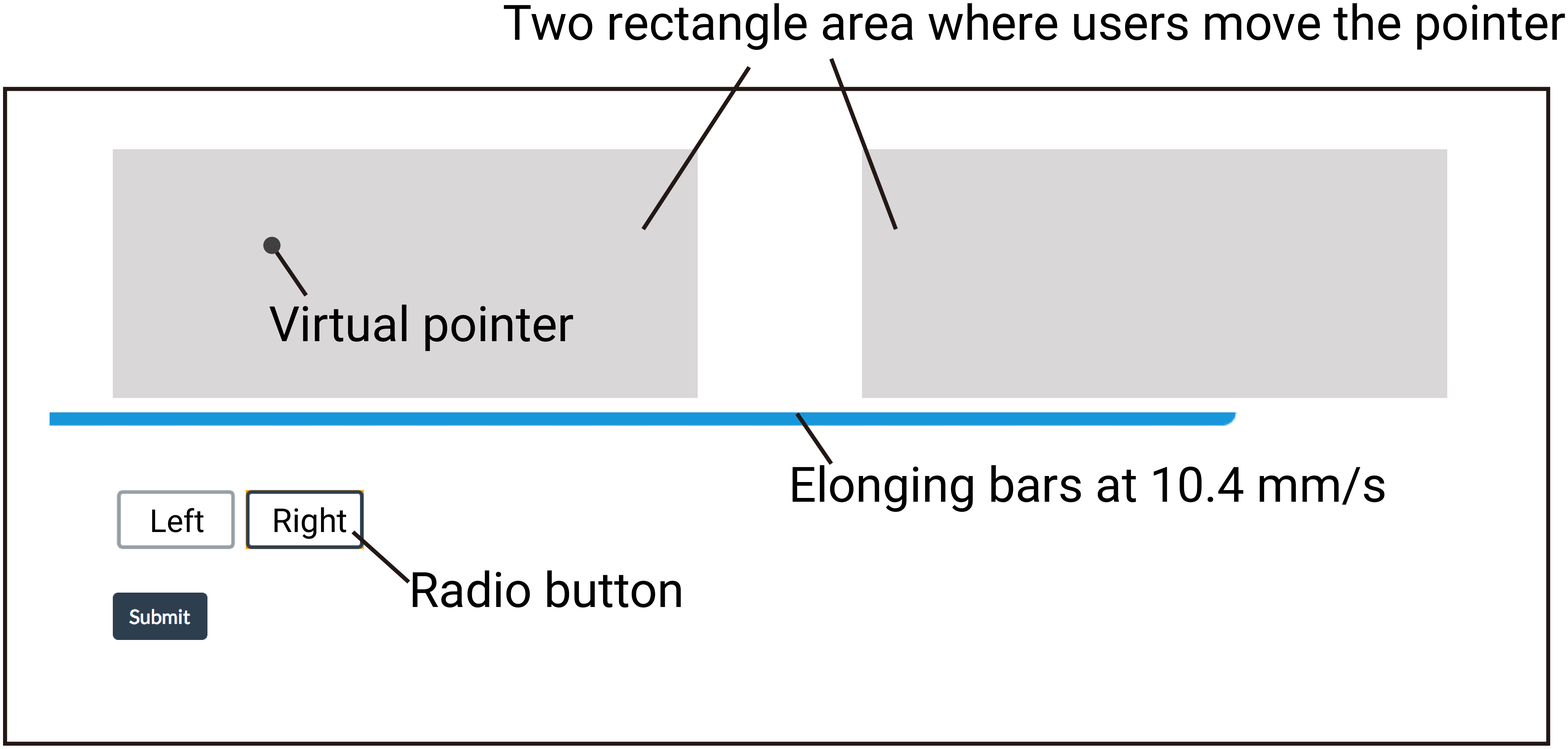}
\caption{Experimental window.}
\label{fig_ex1_window}
\end{figure}

This user study used a within-participants design.
Participants had to move the pen-type device from left to right while watching a virtual pointer in the two different rectangle areas visualized on the screen in succession (Fig.\ref{fig_ex1_window}).

We describe the procedure of one trial in the task.
The participants moved the pen on two rectangular areas from left to right at a constant speed (10.4 mm/s) with their dominant hands.
The system had participants move their pen's speed by illustrating an elongating blue bar on the screen.
The screen visualized a virtual pointer that indicated where they touched, though the position of the pointer was controlled by the pointer movement controller.
Participants were required to grip the pen with their fingers at the position the vibrator was embedded so that they felt the vibration in their fingers.
Participants received vibrational feedback, which in reality was the same in both areas during the movement, although participants were not aware of this.
After participants finishing movement on two areas with the virtual pointer, they stated which area they felt had a rougher vibrotactile texture.
They tapped one of two answer buttons visualized on the screen.

The additional instructions we gave participants were as follows.
We told participants to focus on the vibration while watching the pointer on the screen and asked them to judge the roughness mainly based on the vibrational feedback instead of visual feedback.
We also told them to select one of two buttons randomly if they thought that it was difficult to judge which was rougher.

$\alpha$ of the non-oscillatory area was 0.
$\alpha$ of the oscillatory area was selected from 1.5, 2, 2.5, and 3.
$\lambda$, which was the parameter of the vibrotactile signal generator, was the same in the non-oscillatory and oscillatory area. $\lambda$ was selected from 1/3, 1/5, and 1/7.
Thus, there were four visual conditions and three signal conditions.
The permutation of them was 12 conditions.
Participants performed each condition 10 times.
Thus, each participant conducted 120 trials.
The position of the non-oscillatory was randomly assigned to the left or the right.
The presentation order of these factors was randomly assigned and counterbalanced across participants.

\subsection{Results}

First of all, to determine whether the answers to the question were dissimilar by chance, a chi-square goodness-of-fit test was performed on the frequency of the participants{}' selections under the two rectangle areas against the null hypothesis that the two areas were equally selected (Table \ref{tbl_ex1_result_table}).
Table \ref{tbl_ex1_result_table} shows the number of times the oscillatory area was selected for each condition.
Out of the 100 selections (10 participants multiplied by 10 times) under each condition for the question, the difference was observed to be statistically significant ($p \verb|<| 0.01$) in all conditions.
This shows that when the configuration of $\alpha$ was from 1.5 to 3, participants felt the vibrotactile texture as rougher than when $\alpha$ wass set to zero.
In other words, our proposed method succeeded in modulating roughness perception in all conditions.

\begin{table}[h]
\centering
\caption{The number of answers and chi-square goodness-of-fit values.}
\includegraphics[width=3.5in]{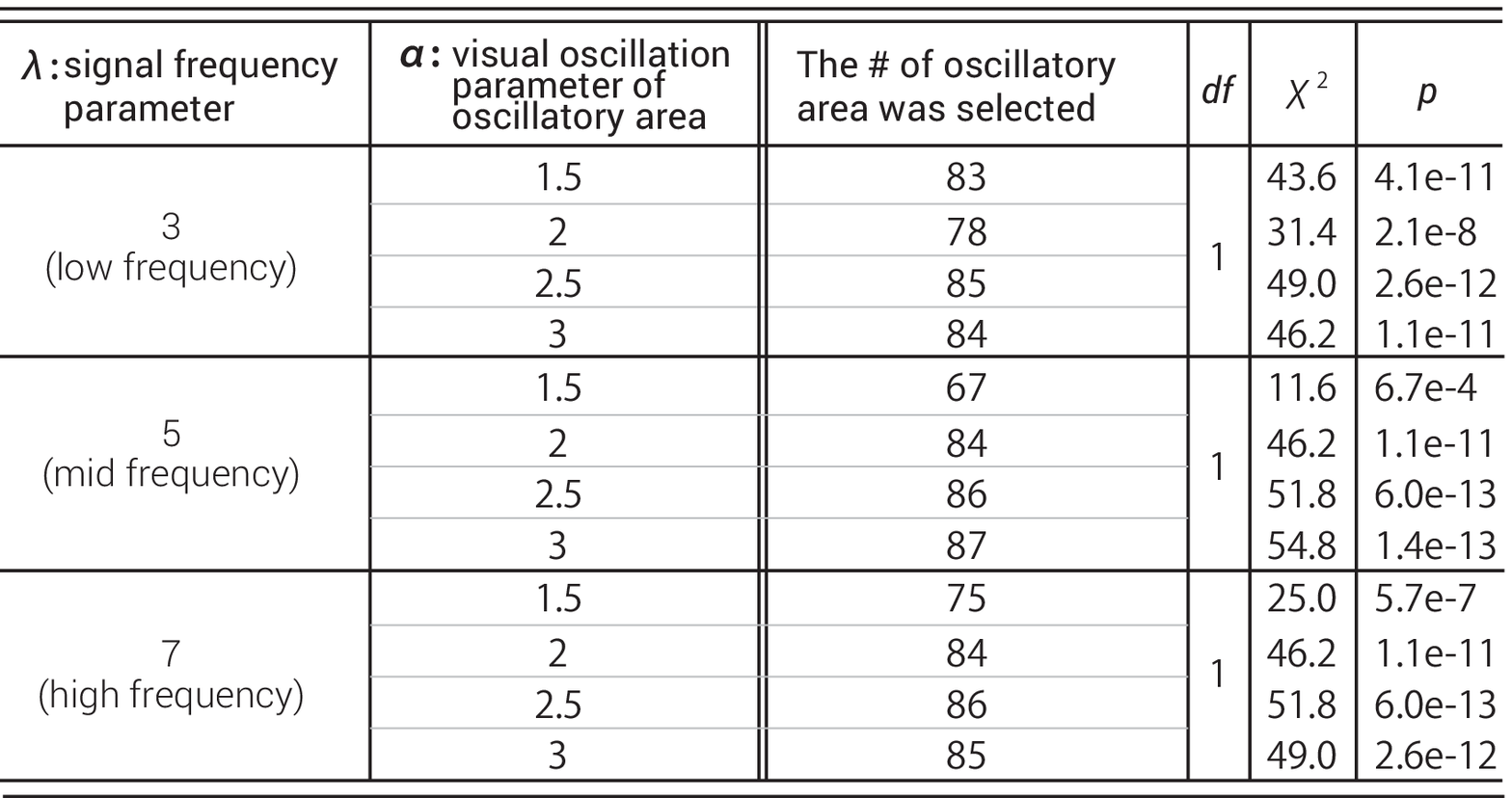}
\label{tbl_ex1_result_table}
\end{table}

Next, to determine whether there is a significant relationship between signal conditions and visual conditions on the frequency of selections, we performed a chi-square test of independence.
The null hypothesis is that signal conditions and visual conditions are independent.
The result showed that there was not significant relationship between them ($ \chi^{2}=1.95, df=6, p=0.92$).
Thus, both conditions are shown to be independent.

\subsection{Discussion}

The results of the chi-square goodness-of-fit test proved hypothesis H1 to be correct under the conditions used in this study.
In other words, users felt the vibrotactile textured surface was rougher by watching visual oscillation of the pointer despite receiving the same vibrational feedback.
Because we asked participants to focus on the vibration while watching the pointer on the screen, participants did not judge roughness only from visual information like in \cite{Watanabe2008}, but they judged it mainly from vibrational feedback.
Our method made users feel the textured surface as rougher at a high probability, which was approximately 80\%.

We have interests in what participants felt about the roughness enhanced by our method.
Was the feeling of roughness different from the feeling of real vibration?
Usually, the haptic feeling induced by pseudo-haptic effect is different from the real haptic feeling.
One participant's comment after all trials suggested a difference in feeling. They said that when they watched visual oscillation, they felt as if they were scanning the rocky surfaces and felt much roughness.
They also said that they felt as if they were scanning the artificial patterned texture when they scanned the surfaces without visual oscillation.

According to the result of the chi-square independence test between factors of the signal condition's parameter and the visual condition's parameter, we can conclude that there was no significant relationship between the parameters.

Our method showed robust effectiveness under the conditions used in this study.
However, we estimate that there will be a range of signal and visual conditions where our proposed method is effective, which was not confirmed in this user study.
Though the pointer visually oscillates back/forth and left/right discontinuously in our method, the visual oscillation of the pointer was normally a few pixels in distance and thus the discontinuity worked.
However, if the size of visual oscillation is too large, the proposed method will break because of the discontinuity.
Participants will think the visual oscillation has nothing to do with vibrations and they cannot imagine that visual oscillation is caused by the virtual fine unevenness, and the illusionary roughness is difficult to modulate.
After the user study, one participant said that when the visual oscillation is too large, they felt a sense of incongruity.
If the size of visual oscillation is too small, participants cannot notice the visual oscillation, and there will be no difference in perceived roughness.
On the other hand, if the frequency of the square wave is too small, the vibration and the visual oscillation does not synchronize and participants would feel a sense of incongruency.
If the frequency of the square wave is too high, the roughness cannot be felt from it and participants would feel a sense of incongruency between the vibration and the visual oscillation.

Based on this consideration, we conclude that we can modulate the roughness perception at the high probability under the visual and signal conditions used in this user study.
In order to finely control perceptual roughness, we consider how much roughness is modulated by our method in another user study, presented in the next section.

\section{User study 2}

We performed two different within-participants studies in User Study 2 comparing six different visual conditions.
User Study 2 was conducted to test hypothesis H2.
In other words, we tested whether participants felt the vibrotactile surfaces as rougher as the size of the visual oscillation of the pointer became larger.

To test this from a quantitative viewpoint as much as possible, we made full use of existing research results on roughness perception.
There are at least two ways in which aspects of vibration might underlie roughness perception.
Roughness might be encoded intensively.
In other words, the intensity of the vibrations determines their perceived roughness, which has been shown in \cite{Miyaoka1999,Hollins2000}.
Alternatively, roughness might be encoded temporally, in which case the frequency composition of the texture-induced vibrations determines the perceived roughness.

Based on the theory stated above, we assumed that users would feel as if the amplitude of vibration is larger or wavelength is larger when they felt vibration as rougher with the proposed method.
Thus, we conducted two different experiments in User Study 2: "amplitude experiment" and "wavelength experiment."
In both experiments, participants compared the roughness of texture between two areas.
We named the two areas as "oscillatory area" and "non-oscillatory area" (Table \ref{fig_ex2_table_condition}).
Visual oscillation was enabled in the oscillatory area, but it was disabled in the non-oscillatory area.
Note that there was vibrotactile feedback in both the oscillatory and non-oscillatory area.

In the "amplitude experiment," participants adjusted the amplitude of the square waveform of the non-oscillatory area until they could feel the roughness of oscillatory and non-oscillatory area perceptually equally.
We expected that participants would adjust the amplitude of the non-oscillatory area larger than the oscillatory area.
If roughness might be encoded intensively and if participants really felt the texture as rougher in the oscillatory area with our method, they would feel as if the amplitude of the vibration would be larger.
For evaluation, we measured the peak-to-peak voltage values on the vibrator and compared those between the oscillatory area and non-oscillatory area.

On the other hand, in the "wavelength experiment," participants adjusted the wavelength of the square waveform in the non-oscillatory area.
We expected that participants would adjust the wavelength of the non-oscillatory area more than in the oscillatory area.
If roughness might be encoded temporally and if participants really felt the texture as rougher in the oscillatory area with our method, they would feel as if the wavelength of the vibration would be larger.

\begin{table}[h]
\centering
\caption{Summary of parameters of vibration and visual oscillation in the amplitude and the wavelength experiments.}
\includegraphics[width=3.5in]{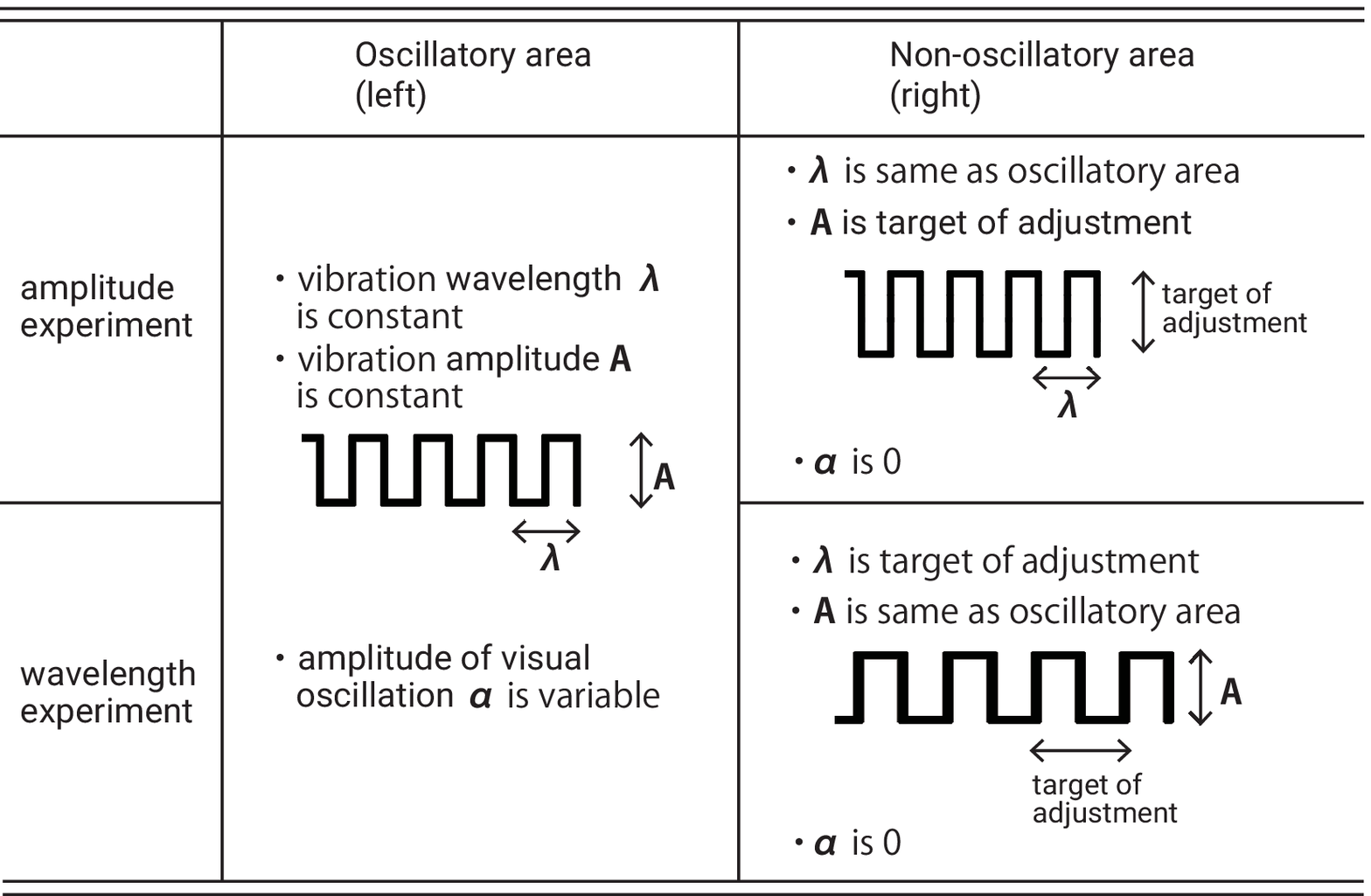}
\label{fig_ex2_table_condition}
\end{table}

\subsection{Participants}
The number of participants was ten (seven males and three females) aged from 22 to 24.
All of the participants were right-handed.
They were screened to determine that they were not depressed or tired because perception can be affected by physical or emotional states.
In the same way as in User Study 1, The University of Tokyo Ethics Committee approved the data acquisition in this paper and written informed consent was obtained from all participants.

\subsection{Experimental system}
The experimental system was the same as in User Study 1.
The system was composed of a touchpad, display, and vibrotactile pen-device.

\subsubsection{Configuration of pointer movement controller}

The pointer position was manipulated in the same ways as in User Study 1 (see Section 3).
The size of visual oscillation $\alpha$ was a variable in this user study.
In this study, we set $\alpha$ of the non-oscillatory area as 0.
We set $\alpha$ of the oscillatory area as 0.5, 1.0, 1.5, 2.0, 2.5, or 3.0.
Also, we refer to the conditions of the oscillatory area that has different $\alpha$ as "visual conditions" in this user study.

\subsubsection{Configuration of vibrotactile signal generator}

In the same way as in User Study 1, we assumed a simply patterned virtual surface.
The virtual textured surface was enhanced with a striped pattern that generates vibrotactile feedback of constant wavelength $\lambda$:

\begin{equation}
y(t) = A\;sgn(\sin (2 \pi \frac{V(t)}{\lambda} + \varphi))
\end{equation}

According to the results of the chi-square independence test in User Study 1, there is no significant relationship between signal and visual conditions.
In this study we would like to investigate the proposed effect by changing many different visual conditions, so we set $\lambda$ of the oscillatory area at 1/5 in this user study.
When the pen's movement speed was 10.4 mm/s (90 pixels/s), the frequency of the square wave was 9 Hz and the wavelength was 1.15 mm/s (10 pixels).
The amplitude $A$ of the oscillatory area was also constant.
This affected the applied voltage on the vibrator.
We measured the peak-to-peak voltage applied on the vibrator when participants moved the pointer in the oscillatory area where $\lambda$ of the oscillatory area was 1/5 and the pen's movement speed was 10.4 mm/s.
The measured value was 4.67 V.

In the "amplitude experiment," the amplitude $A$ of non-oscillatory area was a parameter that participants could manipulate.
In the "wavelength experiment," the wavelength $\lambda$ was a parameter that participants could manipulate.

\subsection{Task design}

\begin{figure}[h]
\centering
\includegraphics[width=3.5in]{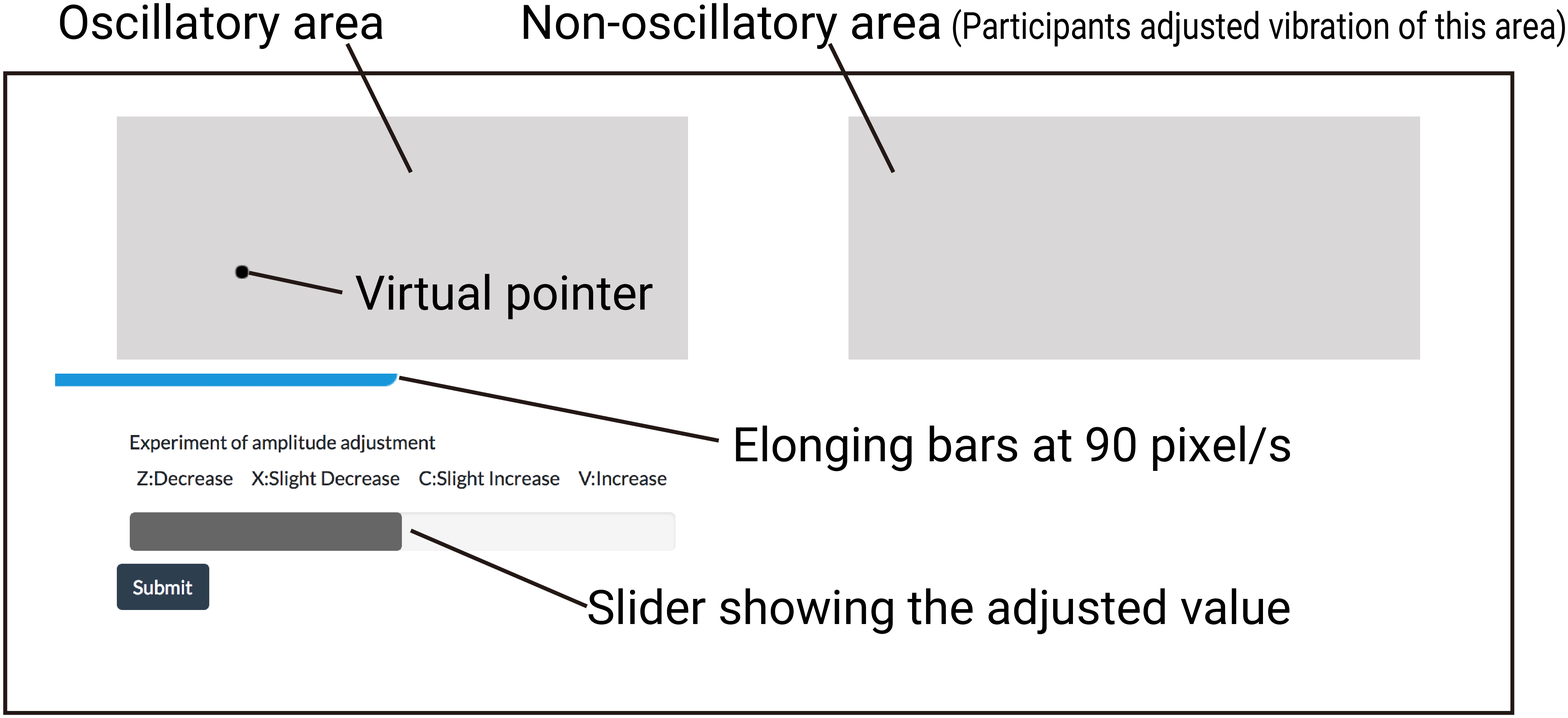}
\caption{Experimental window.}
\label{fig_ex2_window}
\end{figure}

This study used a within-participants design.
Participants had to move the pen-type device from left to right or from right to left watching virtual pointers in the two different rectangular areas visualized on screen in succession (Fig.\ref{fig_ex2_window}).
They could manipulate the parameter of waveforms of vibration on the non-oscillatory area, but the parameter of vibration waveform on the oscillatory area was always constant in this study.
Participants{}' task was to make the subjective roughness of the textured surfaces of both areas the same by adjusting the vibrational parameter of the non-oscillatory area.

We describe the procedure of one trial in the task.
The participants moved the pen in two rectangular areas from left to right or from right to left at the constant speed (1.15 mm/s) with their dominant hands.
The system let participants recognize the speed by illustrating it using elongating blue bars on the screen.
The screen visualized a virtual pointer that indicated where they touched, but the position of the pointer was controlled by the pointer movement controller.
Participants were required to grip the pen with their fingers at the position where a vibrator was embedded so that they felt the vibration with their fingers.
After participants finished movement in two areas with the virtual pointer, they adjusted either amplitude or the wavelength of vibration in the non-oscillatory area by pushing a "decrease", "slight decrease", "slight increase", or "increase" button on the screen.
Initial values of the non-oscillatory area were the same as the oscillatory area, that is, in the amplitude experiment, the initial amplitude $A$ was 4.67 V, and in the wavelength experiment, the initial wavelength $\lambda$ was set at 1/5.

The gray bar on the screen showed the adjusted value as reference information.
The minimum adjustable value was 1/5 times as low as the parameter of the oscillatory area.
The maximum adjustable value was 5 times as high as the parameter of the oscillatory area.
The variables were defined by multiplication of initial value and $10^{(S/100)}$ and the initial value $S = 0$.
When participants increased the parameter, $S$ became larger by 6.
When participants decreased the parameter, $S$ became smaller by 6.
When participants slightly increased the parameter, $S$ became larger by 3.
When participants slightly decreased the parameter, $S$ became smaller by 3.
They adjusted the parameter so that the roughness of both areas felt equal.
They were allowed to move the pen from side to side repeatedly until they finished adjusting.
They moved the pen in whichever area they wanted during adjusting.
There was no time limit for adjustment.

The additional instructions we gave participants were the same as in User Study 1.
We told participants to focus on the vibration while watching the pointer on the screen and asked them to judge the roughness mainly based on the vibrational feedback instead of visual feedback.

The size of visual oscillation $\alpha$ of the non-oscillatory area was selected from 6 values (0.5, 1.0, 1.5, 2, 2.5, and 3), but the size of $\alpha$ of the non-oscillatory area was always 0.
Thus, there were 6 visual conditions for both the amplitude experiment and wavelength experiment.
After participants performed all trials in one of the amplitude experiments or wavelength experiments, they moved on to the other experiment.
Participants performed each condition 5 times.
Thus, each participant conducted 60 trials.
The oscillatory area was on the left, and the non-oscillatory area was on the right.
The presentation order of these factors was randomly assigned and counterbalanced across participants.

\subsection{Results}
\begin{figure}[h]
\centering
\includegraphics[width=3.5in]{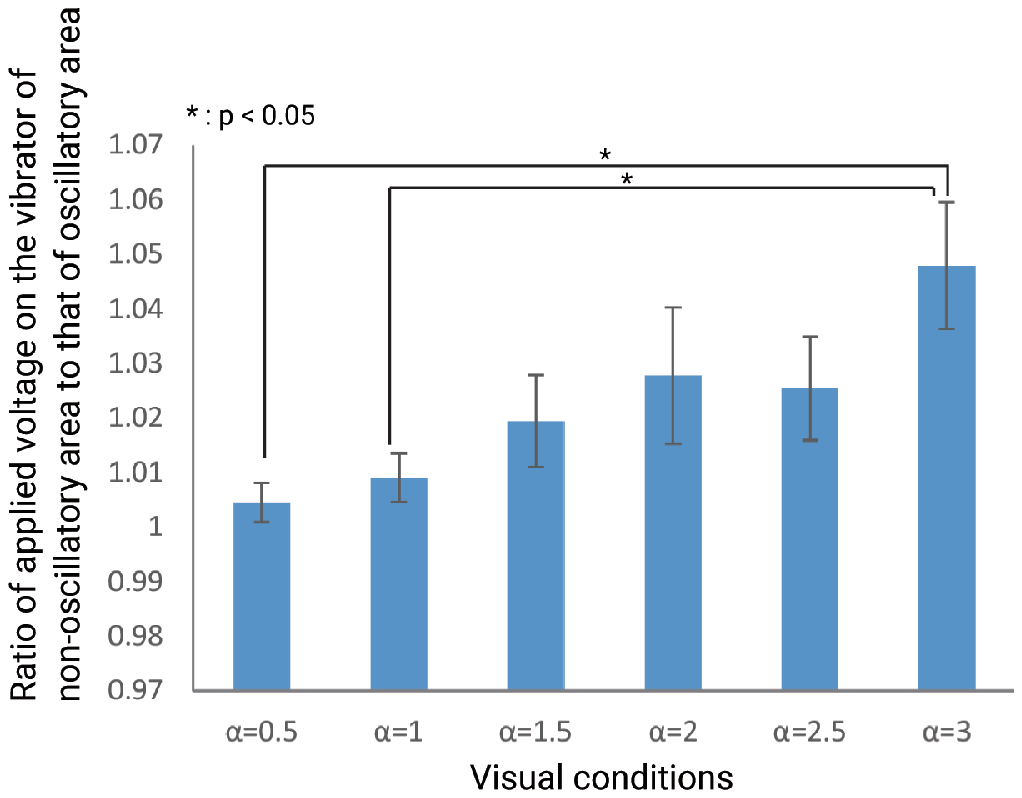}
\caption{The ratio of adjusted vibration voltage of non-oscillatory area to that of oscillatory area. Standard error is also illustrated.}
\label{fig_ex2_result_amp}
\end{figure}

\begin{figure}[h]
\centering
\includegraphics[width=3.5in]{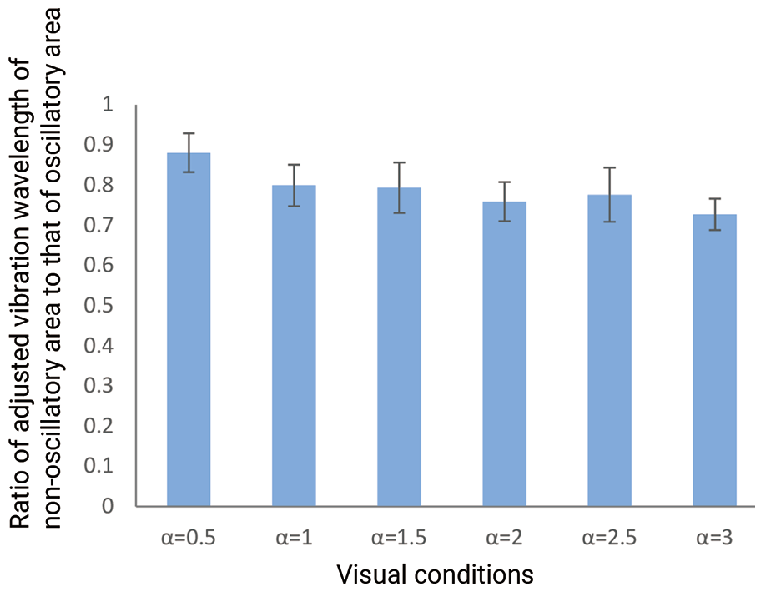}
\caption{The ratio of adjusted vibrational wavelength of non-oscillatory area to that of oscillatory area with respect to visual conditions. Standard error is also illustrated.}
\label{fig_ex2_result_freq}
\end{figure}

Fig.\ref{fig_ex2_result_amp} shows a result of the amplitude experiment.
The horizontal axis represents the visual conditions, and the vertical axis represents the ratio of the adjusted peak-to-peak voltage of the non-oscillatory area to the peak-to-peak voltage of the oscillatory area.
The ratio of voltage means the ratio of amplitude of the square wave presented to the participants between both areas.
Fig.\ref{fig_ex2_result_freq} shows a result of the wavelength experiment.
The horizontal axis represents the visual conditions, and the vertical axis represents the ratio of the adjusted vibrational wavelength of the non-oscillatory area to the wavelength of the oscillatory area with respect to each visual condition.

To determine whether participants felt the vibrotactile surfaces as rougher as the size of the visual oscillating of the pointer was larger, we performed a one-way repeated ANOVA with factors of visual condition's parameter ($\alpha = 0.5, 1.0, 1.5, 2, 2.5, 3$) on the adjusted voltage values for the amplitude experiment and wavelength experiment separately.
We conducted a Shapiro-Wilk test to check normality and a Mauchly's test to check sphericity criteria and proved them in advance of the ANOVA test.
According to the ANOVA results, there was a significant effect of the visual condition on voltage value in the amplitude experiment ($F(5,54) = 2.39, p = 0.019$).
On the other hand, there was no significant effect of the visual condition in the wavelength experiment ($F(5,54) = 2.39, p = 0.47$).

We applied Tukey comparisons for all post-hoc comparisons for the amplitude experiment.
As a result, there was a significant difference under two visual conditions where $\alpha$ was 0.5 and 3.0 ($p \verb|<| 0.05$).
There was also a significant difference under two visual conditions where $\alpha$ was 1.0 and 3.0 ($p \verb|<| 0.05$).

\subsection{Discussions}

ANOVA and Tukey comparisons (Fig.\ref{fig_ex2_result_amp}) show a significant effect of the size of visual oscillation on the perceived amplitude of the vibration.
This shows that as the visual oscillation grew, participants felt the vibration amplitude as larger.
Based on the knowledge of previous studies'{} \cite{Miyaoka1999,Hollins2000} results, which said that the vibration amplitude has an effect on perceived roughness, it is shown that participants perceived the vibrotactile texture as rougher as the visual oscillation grew.
In other words, hypothesis H2 was proved.

In this experiment, initial values of the non-oscillatory area were the same as the oscillatory area, that is, in the “amplitude experiment”, the initial amplitude value was 4.67V and in the “wavelength experiment”, the initial wavelength was set 1/5.
Because the initial values in the experiments were the same, there is a possibility that participants noticed the initial value was the same in both oscillatory and non-oscillatory areas.
In usual experimental settings other than ours, there is a case that whether participants notice the initial values being always the same or not could affects results.
It is because participants can remember how much value they adjust and they are likely to move that volume based on the remembrance.
On the other hand, however in this user study, we assumed that whether participants noticed or not did not affected the results.
It is because we presented 6 different visual conditions in a random order.
Thus participants could not adjust the value from their remembrance and we can conclude that the initial value setting in this study did not affect the results.
Moreover, future studies could involve both ascending and descending series as is commonly done in psychophysical studies \cite{Ehrenstein1999}.

The results suggested that it is possible to control the amount of perceived roughness with our method.
It is expected that we can modulate the perceptual roughness by changing the size of oscillation toward the intended one when there is a mismatch between presented and intended perceptional roughness.
Based on this experiment, we showed an approximate possibility of making surfaces at least 5\% perceptually rougher.

Fig.\ref{fig_ex2_result_freq} shows the results of the wavelength experiment.
We expected before the wavelength experiment that as the visual oscillation size grows, the wavelength grows.
However, the one-way repeated ANOVA showed no significant effect of visual conditions on the perceived wavelength of vibrations.
The reason for this result is assumed to be that we controlled the size of visual oscillation in this study, so the pseudo-haptics affected the vibration's perceived amplitude but did not affect its perceived frequency.
If we control the other variables of visual oscillation, such as quality of randomness and size of oscillation, the perceived frequency might be affected.


\section{General Discussion}


As we analyzed the difference of our method from existing studies on pseudo-haptic in the section of related work, our method is the first research that has made an attempt to modulate the real vibrotactile roughness feeling.
According to the results of the user study 1, it is shown that our method succeeded in making users feel the vibrotactile texture rougher at a high probability under all signal and visual conditions used in the study.
In addition, according to the results of user study 2, the modulation of fine roughness perception can be controlled by changing the size of visual oscillation.
It can be said that our study extended the possibility of pseudo-haptic effect.

Here, we discuss use cases where our method can be applied.
Our method has advantages in terms of contents consistency, that is, it does not need to change the appearance of texture surfaces, in contrast to the studies modifying the texture appearance \cite{Lederman1981, Heller1982}.
Also, our method does not need additional vibrational stimuli in contrast to "signal-based approach" such as \cite{Asano2015,Hollins2000}.
It is expected that our method and "signal-based approach" can be used together to modulate fine roughness perception, but it is a future study.
If only there is a virtual pointer on surfaces and vibrotactile feedback, we can try to apply our method.
Thus, when VR developers would like to use our method in order to modulate the perception of vibrational roughness, they only need to visualize the pointer indicating the position of touch on surfaces.
For the VR applications, we can freely visualize the pointer at the contact area on the surface and thus, our method can be used easily.
For example, one of the assumed VR practical applications is surgical training simulator where trainee holds the instrument in the real world to touch the patient's virtual skin surface and the roughness perception of during the interaction between the instrument and the skin is controlled.
In this study, we tested by user studies under conditions where users manipulate the visualized pointer by pen and we did not test whether the effect would be the same under conditions where users watch the movement of virtual hands or fingertips when they are interacting with virtual surfaces using their hands.
As opposed to VR applications, you may assume that it is difficult to apply our method to touchscreen interaction because apparently usually there is no pointer.
However, there is a possibility of applying it to the touchscreen interaction if there is an alternate icon or an object which works as a pointer.

One of the limitation is that there will be a range of signal and visual conditions where our proposed method is effective, as described in the discussion section in the User Study 1.
As for the visual conditions, though the pointer visually oscillates back/forth and left/right discontinuously in our method, the visual oscillation of the pointer was normally few pixels distances and thus the discontinuity work.
However, if the size of visual oscillation is too large, the proposed method will break because of the discontinuity.
Participants cannot imagine that it is caused by the virtual fine unevenness and thus the illusionary roughness is difficult to modulate.

As for the signal conditions, humans can perceive a larger frequency range of vibrations than those used in this study.
Whether the effect of the proposed method on the vibration of higher frequency is the same should be also tested in the future study.

\section{Conclusion}

This study presents a novel cross-modal modulating method of vibrotactile fine roughness perception.
The user studies yielded the following findings:

\begin{itemize}
 \item When users watched the pointer slightly oscillating back/forth and left/right receiving vibrational feedback, users felt the vibrotactile surfaces more uneven.
 \item The larger size of visual oscillation enlarged the perceived amplitude of the vibrotactile signal wave. The result suggested that we can control the perceived amount of roughness by changing the size of visual oscillation. However, the size of visual oscillation did not affect the perceived frequency of the wave.
\end{itemize}

These results suggests that our method is helpful for modulating vibrotactile perception.
It has a possibility to overcome a mismatch between the perceived vibrotactile feeling and the intended feeling.
As opposed to conventional signal-based approach, our method does not need to modify the vibrotactile signal that is applied to users.
Of course, it is expected that our method can be used in combination with conventional signal-based approach, but it is a future study.

Our method can be used in any scene where users explore virtual surfaces watching the pointer which shows the contact point in the VR applications.
In addition, our approach might be effective even when the pointer is replaced with other representation such as virtual fingertips or tools in order to manipulate the perceived roughness of virtual surfaces, which we will test in the future study.
\acknowledgments{
This research was partially supported by the Ministry of
Education, Science, Sports and Culture, Grant-in-Aid for
Research Activity Start-up, 17H06573, 2017. Yuki Ban is the
corresponding author.
}

\bibliographystyle{abbrv-doi}
\bibliography{bibdata}

\end{document}